\documentclass{aa}  

\usepackage{graphicx}
\usepackage{txfonts}
\usepackage{natbib}
\newcommand{\bz}{\ensuremath{\langle B_z\rangle}}
\newcommand{\bs}{\ensuremath{\langle B \rangle}}

\newcommand{\kms}{\ensuremath{\mathrm{km\,s}^{-1}}}
\newcommand{\vsi}{\ensuremath{v \sin i}}
\newcommand{\te}{\ensuremath{T_{\mathrm{eff}}}}
\newcommand{\llo}{\ensuremath{\log L/L_\odot}}
\newcommand{\bd}{BD-19\,5044L}
\newcommand{\esp}{ESPaDOnS}

\begin{document}

   \title{BD-19 5044L: discovery of a short-period SB2 system with a magnetic Bp primary in the open cluster IC~4725}

   \titlerunning{BD-19 5044L: a magnetic short-period SB2 in IC~4725}

   \author{J. D. Landstreet \inst{1,2}\fnmsep\thanks{Based in part on observations obtained at
       the Canada-France-Hawaii Telescope (CFHT) which is operated by
       the National Research Council of Canada, the Institut National
       des Sciences de l'Univers of the Centre National de la
       Recherche Scientifique of France, and the University of
       Hawaii.} 
     \and O. Kochukhov \inst{3} 
     \and E. Alecian \inst{4,5,6}
     \and J. D. Bailey \inst{7} 
     \and S. Mathis \inst{8,6}
     \and C. Neiner \inst{6} 
     \and G. A. Wade \inst{9}
     \and the BINaMIcS collaboration
}

   \institute{Armagh Observatory, College Hill, Armagh, BT61 9DG, 
              Northern Ireland, UK
              \email{jlandstr@uwo.ca;jls@arm.ac.uk}
         \and
              Department of Physics \& Astronomy, University of 
              Western Ontario, London, Ontario N6A 3K7, Canada
         \and 
              Department of Physics and Astronomy, Uppsala
              University, Box 516, 75120, Uppsala, Sweden 
         \and 
              UJF-Grenoble 1/CNRS-INSU, Institut de Plan\'{e}tologie 
              et d'Astrophysique de Grenoble (IPAG) UMR 5274, 38041, 
              Grenoble, France 
         \and 
              CNRS-IPAG, F-38000, Grenoble, France 
         \and 
              LESIA, Observatoire de Paris, PSL Research University, 
              CNRS, Sorbonne Universit\'es, UPMC Univ. Paris 06,  
              Univ. Paris Diderot, Sorbonne Paris Cit\'e, 
              5 place Jules Janssen, 92195 Meudon, France
         \and 
              Max Planck Insitut f\"{u}r Extraterrestrische Physik, 
              Giessenbachstrasse 1, 85748 Garching, Germany
         \and 
              Laboratoire AIM Paris-Saclay, CEA/DRF, CNRS, 
              Universit\'{e} Paris Diderot, IRFU/SAp Centre de Saclay,
              F-91191 Gif-sur-Yvette, France 
         \and 
              Department of Physics, Royal Military College of
              Canada, PO Box 17000, Stn Forces, Kingston, Ontario 
              K7K 7B4, Canada
}

   \date{Received September 15, 1996; accepted March 16, 1997}

 
  \abstract
   {Until recently almost nothing was known about the evolution
   of magnetic fields found in upper main sequence Ap/Bp stars during
   their long main sequence lifetime. We are thus studying magnetic
   Ap/Bp stars in open clusters in order to obtain observational
   evidence of how the properties of Ap/Bp magnetic stars, such as
   field strength and structure, evolve with age during the main
   sequence. One important aspect of this study is to search for the
   very rare examples of hot magnetic stars in short-period binary
   systems among magnetic cluster members. }
   {In this paper we characterize the object BD-19~5044L, which is
   both a member of the open cluster IC~4725 = M~25, and a short-period SB2
   system containing a magnetic primary star.}
   {We have obtained a series of intensity and circular polarisation
   spectra distributed through the orbital and rotation cycles of
   BD-19~5044L with the ESPaDOnS spectropolarimeter at CFHT. From
   these data we determine the orbital and stellar properties of each
   component. }
   {We find that the orbit of BD-19~5044L\,AB is quite eccentric ($e =
   0.477$), with a period of 17.63~d. The primary is a magnetic Bp
   star with a variable longitudinal magnetic field, a polar field strength of
   $\sim 1400$~G and a low obliquity, while the secondary is probably
   a hot Am star and does not appear to be magnetic. The rotation
   period of the primary (5.04~d) is not synchronised with the orbit,
   but the rotation angular velocity is close to being synchronised
   with the orbital angular velocity of the secondary at periastron,
   perhaps as a result of tidal interactions. Because this system is a
   member of IC~4725, the two stars have a common age of $\log t =
   8.02 \pm 0.05$~dex. }
   {The periastron separation is small enough (about 12 times the
   radius of the primary star) that BD-19\,5044L may be one of the
   very rare known cases of a tidally interacting SB2 binary system
   containing a magnetic Ap/Bp star.}


   \maketitle
%

\section{Introduction}

%


Magnetic fields have been found in a small percentage of upper main
sequence stars (mostly chemically peculiar or Ap/Bp stars) since the
1940s. It is possible to detect quite small fields, down to mean
longitudinal field values \bz\ of a few G, and to establish the
surface magnetic field geometry and to map the anomalous and patchy
atmospheric chemical abundances that often accompany magnetism
\citep[see for example reviews by][]{MestLand05,DonaLand09,BagnLand15}. 

However, although a lot is now known about magnetism in field stars
near the Sun, it has not been possible from these data to establish
clearly how the observed fields and accompanying chemical
peculiarities evolve during the long main sequence lifetime, because
it has been very difficult to obtain accurate ages for field stars.
In the past decade, major improvements in the sensitivity and
wavelength coverage of spectropolarimeters have made it possible to
search for and study magnetic fields in considerably fainter Ap/Bp
stars than in the past, and it has become practical to study magnetic
stars in open clusters of known age. Data for magnetic stars in
clusters of various ages have established clearly that the surface
magnetic fields of upper main sequence late B stars are largest
(typically several kG) early in the main sequence life, and that these
fields decline quite strongly during the main sequence, so that almost
all magnetic stars near the terminal-age main sequence have relatively
weak surface fields, typically a factor of ten or more smaller than
their values close to the zero-age main sequence
\citep{Landetal08}.

The sample of cluster magnetic stars used to study the evolution of
surface field strength is made up largely of stars that have not been
studied very much until recently, as many are fainter than $m_V =
10$. The observations of \citet{Bagnetal06} and \citet{Landetal08}
were the first magnetic measurements made of most of them. In order to
complete a first survey of a statistically interesting sample of stars
in a reasonable amount of time, early magnetic measurements of the
peculiar open cluster stars were mostly snapshots of one or two
measurements.

Following the early survey, a second stage of investigation currently
in progress is to study in detail a fraction of the open cluster
magnetic stars, and to map their field geometries and surface
abundances, in order to discover how these evolve as a function of
main sequence age. This has led to observing programmes obtaining a
number of observations per star, which in turn has led (among other
things) to the discovery of a small number of open cluster
spectroscopic binary systems containing a magnetic Ap/Bp star.

One such object has turned out to be extremely interesting. The Bp object
\bd\ in the open cluster IC~4725 ($\log {\rm age} \approx 8.02$) 
\citep{Landetal07} is shown to be a double-lined spectroscopic binary 
made up of two upper main sequence stars \citep[as suspected
by][]{Paunetal11}. The more massive of the two stars shows a clear
magnetic field, while the less massive star is apparently non-magnetic. We
find the orbital period of the system to be 17.6\,day, with rather
high eccentricity, suggesting that the two stars may have interacted
significantly during their main sequence lifetimes. Such close
binaries containing a magnetic upper main sequence star are very rare
\citep{Carretal02,Alecetal15,Neinetal15a,Neinetal15b}, and so this 
system is well worth studying in detail. 

In this paper we report our extensive spectropolarimetric observations
of \bd\,AB, which we analyse and model to provide a first detailed
description of this system.

\section{Observations}





\begin{figure}[ht]
  \scalebox{0.35}{\includegraphics*{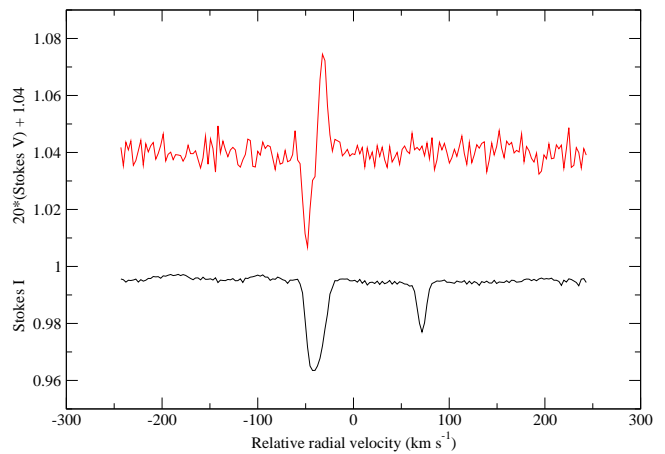}}
  \caption{\label{Fig_sample_lsd_plot} An example of an LSD plot, for
  the spectrum from CFHT file 1822453pn.s. The lower curve is the mean
  Stokes $I$ spectral line, with the primary star at $-45$\,\kms\ and
  the secondary at $+70$\,\kms. The lack of symmetry about the line
  centre of the mean primary line is clearly visible, while the
  secondary line is both constant in shape and symmetric. The upper
  curve is the mean Stokes $V$ circular polarisation, multiplied by
  $20$ and shifted upward to $+1.04$ for ease of comparison. This
  Stokes component clearly shows that a field is present in the
  primary, while no evidence of a field is seen in the secondary. }
\end{figure}

The object \bd\ = CPD-19~6897 is a probable member of the open cluster
IC~4725 = Messier~25, a small cluster at about 620~pc from the
Sun \citep{Kharetal05}. \bd\ has $V = 10.18$ and, based on the single-star photometric
calibration of available $uvby$ photometry, has an effective
temperature of about $\te = 12800$\,K.  No high-precision astrometry is
available, but membership is supported by the good agreement of
stellar effective temperature and luminosity
\llo\ with evolution tracks fitted to other cluster stars
\citep{Bagnetal06,Landetal07}. A further test of membership is to compare the
systemic mean radial velocity $V_0 = 5.30 \pm 0.18$\,\kms of \bd\ (see
below) with the cluster mean radial velocity. The situation in the
literature concerning the radial velocity of IC~4725 is quite
confused, but we may compare the mean radial velocity of \bd\ directly
with our own (unpublished) measured radial velocities of two other
cluster stars, HD\,170836 ($V_{\rm R} = 3.80 \pm 0.5$\,\kms) and
HD\,170860 ($V_{\rm R} = 6.20 \pm 0.5$\,\kms); the close agreement
supports cluster membership (and suggests a cluster mean velocity
close to $+5$\,\kms).

Photometric and spectral classification \citep{Mait85,Paunetal11}
agree that the star is chemically peculiar, but a first effort in 2004
to measure the field using ESO's FORS1 low-resolution
spectropolarimeter did not detect a magnetic field
\citep{Bagnetal06}. However, two further field measurements with the
\esp\ spectropolarimeter at the Canada-France-Hawaii Telescope (CFHT)
successfully detected a mean longitudinal field of $\bz \approx -250
\pm 40$\,G. The field strength was evaluated using the method of
Least-Squares Deconvolution \citep[LSD][]{Donaetal97} as discussed
by \citet{Landetal08}.  The field measurements were obtained on two
successive nights. On the first night the mean Stokes $I$ LSD line profile was
fairly symmetric, while on the second it appeared quite lumpy, which
is not uncommon for magnetic stars with highly non-uniform surface
abundances. Neither spectrum strongly suggested that the object is a
binary star system.

When the campaign to obtain rotationally phased field measurements was
started on the CFHT in semester 2015A, \bd\ was included in the
programme. On the first night of 2015A observations (MJD 57200.320),
the star showed a clearly separated SB2 spectrum, with a magnetic
field detected in the primary star only. As no rotation or other
period was known for this system, it was observed on ten nights in
2015A and six more nights in 2015B, after which \bd\ moved too close
to the Sun for further observation. Two more \esp\ observations were
obtained during semester 2016A, and three during 2016B.

The observations reported here were all made with the \esp\
spectropolarimeter. This instrument obtains Stokes (intensity or flux)
$I$ and (circularly polarised) $V$ spectra covering almost all of the
spectral window between 3800\,\AA\ and 1.04~$\mu$m with spectral
resolution of about 65\,000. The spectra were typically exposed for
about an hour, and signal-to-noise (S/N) ratios of about 200 per pixel
were achieved near 5000\,\AA. The circularly polarised spectra acquired
were reduced at the CFHT using Libre-Esprit, a dedicated software package
based on Esprit \citep{Donaetal97}.  We retrieved the combined
polarised spectra, both normalised and non-normalised. The presence of
the two components of the SB2, separated by some tens of \kms, were clearly
visible in most of the spectra. Both stars are quite sharp-lined. No
signature of a magnetic field was visible in the $V$ spectra of
individual spectral lines.

For each spectrum we then created a mean of all significant spectral
lines and a similar mean of the circular polarisation spectrum over
the same lines, using the LSD technique. This produced mean
line and polarisation profiles with S/N of the order of 2000. An
example of such a profile is shown in
Fig.~\ref{Fig_sample_lsd_plot}. We also created for each mean spectrum
another mean polarisation spectrum with the subexposures added in such
a way as to cancel out any real polarisation signal (a null or $N$
spectrum), in order to look for spurious instrumental polarisation
signals. In every case the $N$ spectrum shows nothing but photon
noise.

The LSD plots gave us mean stellar absorption lines whose radial
velocities could be measured quite accurately, and revealed in every
spectrum the clear S-shaped signature of a magnetic field, always of
approximately the same shape, and always associated with the spectrum
of the primary star. Thus this SB2 system is composed of a brighter
star with a magnetic field, and a fainter one that shows no sign at
all of magnetism.

The radial velocities of both stellar components of the system were
measured using the ``e'' (equivalent width, integrated flux, and
centre) function of the IRAF\footnote{IRAF is distributed by the
National Optical Astronomy Observatories, which is operated by the
Association of Universities for Research in Astronomy (AURA), Inc.,
under cooperative agreement with the National Science Foundation.} 
``splot'' command. Repeated radial velocity measurements suggest that
(except when the two lines overlap) the internal precision of the
measurements is about $\pm 0.5$\,\kms. The scatter observed in repeated
measurements near the same orbital phase in different orbits (see
below, Sect. 3) suggests that this precision estimate is realistic for
the secondary (except when the two spectra begin to overlap). However,
the LSD profiles of the primary star are variable in shape and
equivalent width, a normal situation for a magnetic Ap or Bp star. As
a result, repeated velocity measurements at the same orbital phase
(see below) do not repeat to better than about $\pm 2$\,\kms. This
strongly suggests that the rotation period of the magnetic primary
star is not yet synchronised with the orbital period.

The mean longitudinal magnetic field \bz\ (the mean line-of-sight field
component, averaged over the visible hemisphere) was also
estimated from the LSD spectra, using the expression 
\begin{equation}
   \bz = -2.14\,10^{12} \frac{\int v V(v) dv}
                            {z \lambda_0 c \int (I_{\rm cont} - I(v))dv},
\end{equation}
where the intensity $I$ and circular polarisation $V$ are expressed as
function of velocity $v$ relative to the heliocentric standard of
rest, $I_{\rm cont}$ is the continuum intensity, $\lambda_0$ is the
mean wavelength of the lines contributing to the LSD mean (in \AA),
$z$ is the mean Land\'{e} factor of the lines, and $c$ is the speed of
light \citep{Math89}. Because this measurement is mathematically
equivalent to the measurement of the centroid separation of the two
circularly polarised components of the mean spectral line, no
correction needs to be applied to the measurements because of dilution
of the spectrum by the unpolarised secondary spectrum when the two
stellar lines are clearly separated. No field is detected in the
secondary star with measurement uncertainty of typically 100\,G.

The measured radial velocities of the two components and the measured
values of \bz\ of the primary star, starting with the first two
measurements from 2008, are listed in Table~\ref{Tab_rv_bz}. Note that
\bz\ was not measured for phases at which the primary and
secondary LSD lines are blended with one another.

\begin{table}[th]
\begin{center}
  \caption{CFHT exposure identification (odometer) number, Modified
    Julian Date at mid-exposure, radial velocities of primary and
    secondary ($V_{\rm R1}$ and $V_{\rm R2}$ respectively), and
    measured longitudinal field \bz\ of the primary star with 
    uncertainty $\sigma_{Bz}$.
    }
\label{Tab_rv_bz}
\begin{tabular}{ccccccc }\hline 
Spectrum &  MJD     & $V_{\rm R1}$ & $V_{\rm R2}$ &  \bz    \\
ID number &         &  (\kms)      &  (\kms)      &  (gauss)\\
\hline
1005093  &  54650.448  & $ (4.) $  & $ (6.) $  & $             $ \\   
1005265  &  54651.288  & $ -2.85$  & $ 17.00$  & $             $ \\
1816872  &  57200.320  & $ 23.36$  & $-21.43$  & $ -201 \pm  46$ \\
1817069  &  57201.364  & $ 21.08$  & $-21.14$  & $ -362 \pm  37$ \\
1817272  &  57202.418  & $ 20.53$  & $-19.48$  & $ -416 \pm  34$ \\
1817691  &  57203.466  & $ 20.92$  & $-16.64$  & $ -425 \pm  33$ \\
1818026  &  57204.431  & $ 18.58$  & $-12.49$  & $ -204 \pm  27$ \\ 
1818214  &  57205.431  & $ 13.69$  & $ -6.66$  & $ -229 \pm  22$ \\ 
1818709  &  57207.445  & $ (0.) $  & $ (8.) $  & $             $ \\
1822070  &  57225.353  & $ -3.06$  & $ 19.99$  & $ -230 \pm  32$ \\  
1822241  &  57226.336  & $-20.41$  & $ 42.66$  & $ -271 \pm  36$ \\
1822453  &  57227.318  & $-39.64$  & $ 71.43$  & $ -409 \pm  29$ \\
1823931  &  57233.295  & $ 20.16$  & $-17.32$  & $ -434 \pm  29$ \\
1834183  &  57285.244  & $ 18.69$  & $-12.74$  & $ -227 \pm  80$ \\
1834731  &  57289.233  & $ 24.37$  & $-21.40$  & $ -338 \pm  66$ \\
1834735  &  57289.275  & $ 24.36$  & $-21.10$  & $ -319 \pm  45$ \\
1835709  &  57296.226  & $ -8.58$  & $ 26.98$  & $ -258 \pm  50$ \\
1835937  &  57297.227  & $-28.29$  & $ 53.61$  & $ -372 \pm  45$ \\
1844483  &  57324.231  & $ 23.54$  & $-21.21$  & $ -378 \pm  37$ \\
1915882  &  57474.628  & $-44.25$  & $ 78.56$  & $ -527 \pm  70$ \\ 
1941922  &  57563.439  & $-34.83$  & $ 66.45$  & $ -192 \pm  75$ \\  
1976630  &  57613.439  & $ -6.04$  & $ 24.59$  & $ -206 \pm  66$ \\
1976987  &  57615.272  & $-41.27$  & $ 73.56$  & $ -416 \pm  55$ \\
1977150  &  57616.234  & $-37.72$  & $ 69.49$  & $ -374 \pm  42$ \\

\hline
\end{tabular}
\tablefoot{Values in parentheses are derived from blended lines and are
    considerably more uncertain than those obtained when the two
    spectra are well separated. The RMS dispersion of primary lines
    around the mean RV curve is 0.9\,\kms, and that of the secondary 
    is 1.4\,\kms. }
\end{center}
\end{table}

\section{Orbit of the SB2}




\begin{figure}[ht]
\scalebox{0.35}{\includegraphics*{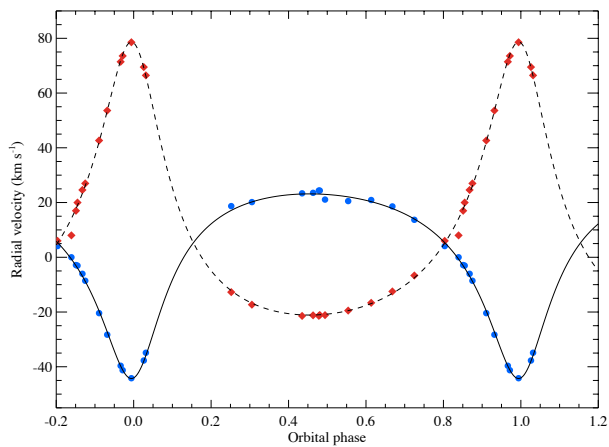}}
\caption{\label{Fig_velocity_curve} Radial velocity curves as a
  function of orbital phase measured relative to the velocity extremum, 
  using the best period of 17.63011~d. The
  velocities of the primary are plotted with blue circles, those of the
  secondary as red diamonds.  Smooth curves are from the model
  discussed in the text.}
\end{figure}

\begin{table}[th]
\begin{center}
  \caption{Orbital elements and uncertainties of the SB2 system \bd\
  computed as described in the text}
\label{Tab_sb2_params}
\begin{tabular}{ccccc}\hline 
Parameter &  Value         & Uncertainty & units \\
\hline

$RMS(1)$  &      0.91      &             & \kms \\
$RMS(2)$  &      1.40      &             & \kms \\
$P_{\rm orb}$ & 17.63011   &   0.00035   & days \\
$T_0$ (MJD) & 54653.915    &   0.070     & days \\
  $K_1$   &     33.73      &   0.36      & \kms \\
  $K_2$   &     49.88      &   0.38      & \kms \\
  $V_0$   &      5.30      &   0.19      & \kms \\
  $e$     &      0.4738    &   0.0063 \\
$\omega$  &    186.7       &   1.0       & deg \\
$M_1/M_2$ &      1.479     &   0.019 \\
$a_1 \sin i$ &  10.35      &   0.12      & $R_\odot$ \\
$a_2 \sin i$ &  15.31      &   0.13      & $R_\odot$ \\
$M_1 \sin^3 i$ & 0.4350    &   0.0081    & $M_\odot$ \\
$M_2 \sin^3 i$ & 0.2942    &   0.0059    & $M_\odot$ \\

\hline
\end{tabular}
\end{center}
\end{table}

Seventeen spectra were obtained in 2015 spanning about four months
with a variety of time intervals between spectra. When the radial
velocity data from 2015 are plotted against time, it is clear from the
data on MJD~57207 and 57225 (when the radial velocities of the two
components of the SB2 system are approximately equal and the velocity
of the primary is becoming negative) that the orbital period is
approximately 17 or 18~d.  Plotting all 2015 data on periods around
these values, and requiring smooth variation of the orbital
velocities, the period was quickly refined to about $17.62 \pm
0.04$~d. We then considered the two data points from 2008, on
MJD~54650 and 54651.  Fortunately these two data points, by chance,
identify the same velocity crossover phase as the two dates mentioned
above, and limit the allowed periods to 17.510, 17.631, and 17.753~d.
With the restrictions derived from the 2015 data alone, we find an
orbital period of $17.630 \pm 0.005$\,d. Adding five new \esp\ radial
velocities from 2016, the final best fit orbital period becomes $P_{\rm orb} =
17.63011 \pm 0.00035$\,d.

The radial velocity curves have been modelled in detail using the
Liege Orbital Solution Package \citep[LOSP; ][]{Sanaetal06}, a set of
FORTRAN programmes that fit the radial velocity data and provide model
values of the important parameters of the SB2 system. They were also
modelled using a programme created by one of us (O. Kochukhov). This
code, written in IDL, performs non-linear least-squares fitting of the
orbital radial velocity variationsbased on the Astrolib {\tt helio\_rv}
routine.  Both programmes provided very similar orbital solutions. As
the orbit computed with Kochukhov's programme fits the observations
somewhat better than the orbit produced by LOSP, we have adopted the
orbital solution found by this programme.

The full set of radial velocities is plotted in
Fig.~\ref{Fig_velocity_curve}, using $T_0 = 54653.915$ as origin. It is
clear from the figure that the adopted period describes the radial
velocity variations very well, but that there is a significant phase
gap of about 0.20 in the available radial velocity data. However, this
missing phase interval does not introduce significant uncertainties
into the analysis of the orbit of the SB2 system. The fit to the two
sets of radial velocities generated by Kochukhov's programme are also
shown in the Figure.

The SB2 system parameters derived from the binary orbit fit are listed in
Table~\ref{Tab_sb2_params}. In this table, the $RMS$ values are the
dispersions of the individual data points around the fitted orbital
solution, $P_{\rm orb}$ is the orbital period, $T_0$ is an MJD of maximum
velocity difference, $K_1$ and $K_2$ are the velocity semi-amplitudes
of the two stars, $V_0$ is the system velocity, $e$ is the orbital
eccentricity, $\omega$ is argument of perihelion, $M_1$ and $M_2$ are
the masses of the primary and secondary stars, $a_1$ and $a_2$  are
the orbital semi-major axes of the individual stars around their
common centre of mass, and $i$ is the orbital inclination relative to
the plane of the sky.  (Subscripts 1 refer to component A, subscripts
2 to component B.)

The dispersion of radial velocities about the best-fit orbit allows us
to estimate the total uncertainties of our radial velocity
measurements as $RMS(A) = 0.9$ and $RMS(B) = 1.4$\,\kms. These values
are somewhat larger than the internal measurement uncertainties
estimated above, but presumably include such additional effects as
physical scatter in the line positions (e.g. due to spectrum
variability of the primary) and small wavelength calibration
uncertainties in the spectra.

Note that the radial velocities of the primary star show an apparent
jump near phase 0.50. This is probably a symptom of intrinsic
variability of the primary spectrum, together with the lack of
synchronism of the orbital and rotational periods of the primary, as
mentioned in the previous section.

\section{Magnetic field of the Bp star}




\begin{figure}[ht]
\scalebox{0.37}{\includegraphics*{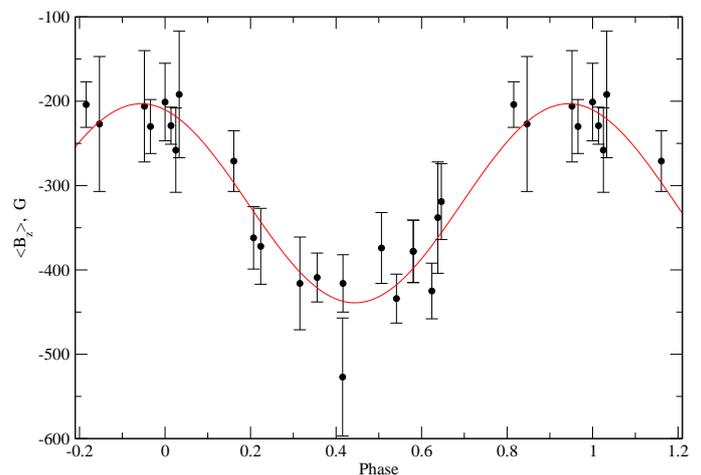}}
\caption{\label{Fig_bz_vars} Variations of the magnetic moment \bz\ of
  \bd\ A phased with a rotational period of 5.041~d. The origin of the
  rotational phase system is the first \bz\ measurement in 2015, from
  Table~\ref{Tab_rv_bz}. The smooth curve is the best sine wave fit to the
  data. }
\end{figure}

\begin{figure}[ht]
\scalebox{0.37}{\includegraphics*{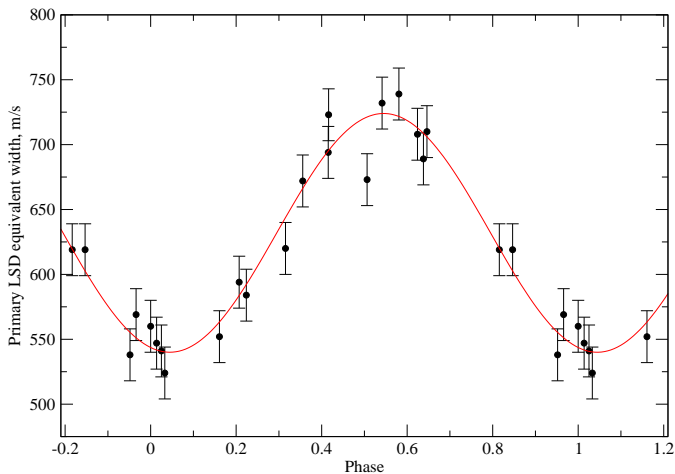}}
\caption{\label{Fig_wlambda_vars} Variations of equivalent width of the LSD
  line of the primary star of \bd, phased with a rotational period of
  5.041~d. The origin of the phase system is the first \bz\
  measurement in 2015, from Table~\ref{Tab_rv_bz}. The smooth curve is
  the best sine wave fit to the data. }
\end{figure}

Table~\ref{Tab_rv_bz} shows that the magnetic field of the primary
star never changes sign, and it is not immediately obvious that the
apparent mild variations are significant. In fact the dispersion of
the \bz\ values is about 85~G, $2\times$ larger than the median \bz\
uncertainty. We have tested these data for (rotational) periodic
variability with a periodogram programme, {\sc dchi.f}, that was used to fit a sine
wave for all significantly different periods between 100 and 1~d by
minimising the reduced chi-squared statistic of the best fit for each
of the many frequencies. The resulting chi-squared fit spectrum,
somewhat surprisingly, shows only two significant periods: $5.040 \pm
0.026$, and its one-day alias $1.2433 \pm 0.0023$~d. The best fit with
the longer period has $\chi^2/\nu = 0.61$, while for the shorter
period the best value is 1.22. Thus both periods fit the data quite
acceptably. In view of the obviously small \vsi\ seen in
Fig.~\ref{Fig_sample_lsd_plot} (about 14\,\kms, see next section),
which would allow a rotation period as long as about 2~wk), the
shorter period is rather unlikely, so this results strongly suggests a
rotation period of 5.04~d. The variation of the magnetic field on this
period is shown in Fig.~\ref{Fig_bz_vars}.

We have tested this period by computing the sine wave chi-squared fit
spectrum for the equivalent width of the primary star's LSD line,
which is found to vary between about 0.54 and 0.74\,\kms, with
uncertainties of about 0.02\,\kms. This second periodogram reveals
exactly the same periods, $5.041 \pm 0.011$~d and $1.2435 \pm
0.0007$~d. The periods determined from equivalent width variations are
more precise than those from \bz\ variations because the full
amplitude of equivalent widths variations is about $8 \sigma$
while that of \bz\ measurements is only about $5.5 \sigma$, where
$\sigma$ is a typical measurement uncertainty. The
minimum values of $\chi^2/\nu = 0.79$ for the 5\,d period while that of
the 1.24\,d period is 1.66. This difference favours the longer period
but is not conclusive. However, on physical grounds we again prefer
the longer period. The variations of equivalent width of the primary
LSD line is shown in Fig.~\ref{Fig_wlambda_vars}. (The small value of
the $\chi^2/\nu$ statistic, and the obviously small scatter around the
mean curve in Fig.~\ref{Fig_bz_vars}, suggest that our LSD \bz\
measurements have rather conservative uncertainties.)

We conclude that the rotation period of \bd\,A is very probably $P_1 =
5.041 \pm 0.011$~d. It is notable that the rotation period is about
3.5 times shorter than the orbital period. The system is clearly
widely enough separated that synchronism of orbital and rotational motions
has not been achieved in the $1.0\,10^8$\,yr since the cluster (and
presumably the binary system) formed. (See the discussion of this
issue in Sect.\,6.)

It is obvious from the plotted variations of \bz\ and equivalent width
with phase on the 5.041~d period that the largest value of the
primary's equivalent width occurs close to the largest absolute value of
\bz, or at the point where the line of sight to the star passes
closest to the visible magnetic pole, while the equivalent width is
smallest where the line of sight is near the magnetic
equator. The LSD line is dominated by even-Z iron peak elements,
particularly by Fe, and this result suggests that the iron abundance
on this star increases from the equator towards the visible magnetic
pole. A visual comparison of the equivalent widths of a number of
lines of He, Mg, Si, Ti, Cr and Fe, made by overplotting a primary
star spectrum obtained near minimum line strength (magnetic phase 0.0)
with one obtained near maximum line strength (magnetic phase 0.5)
shows that He and Mg vary rather little, while the lines of Si and the
three iron peak elements all vary together in the same sense as the
LSD line. It appears that on the surface of \bd\ the elements Si, Ti,
Cr and Fe all have higher abundance near the magnetic pole than near
the equator. This provisional conclusion should be confirmed by
detailed mapping, but in any case it is clear that the abundances of
several spectroscopically important elements vary strongly over
the stellar surface.

Comparing Fig~\ref{Fig_bz_vars} to Fig~\ref{Fig_wlambda_vars}, it is
also clear that there is a small phase shift between the extrema of
about 0.1 cycle -- the largest absolute value of \bz\ occurs about 0.1
cycle before maximum equivalent width of the LSD spectral line. This
situation clearly points to a distribution of abundances over the
visible surface of \bd\,A that is not cylindrically symmetic around
the magnetic field axis of symmetry (if indeed the field itelf is
approximately axisymmetric, which is unknown). This situation is
commonly observed in detailed mapping studies
\citep[e.g.][]{KochWade10,Kochetal11}.

\section{Physical properties of the component stars of the system}


\begin{figure*}[ht]
\scalebox{0.72}{\includegraphics*{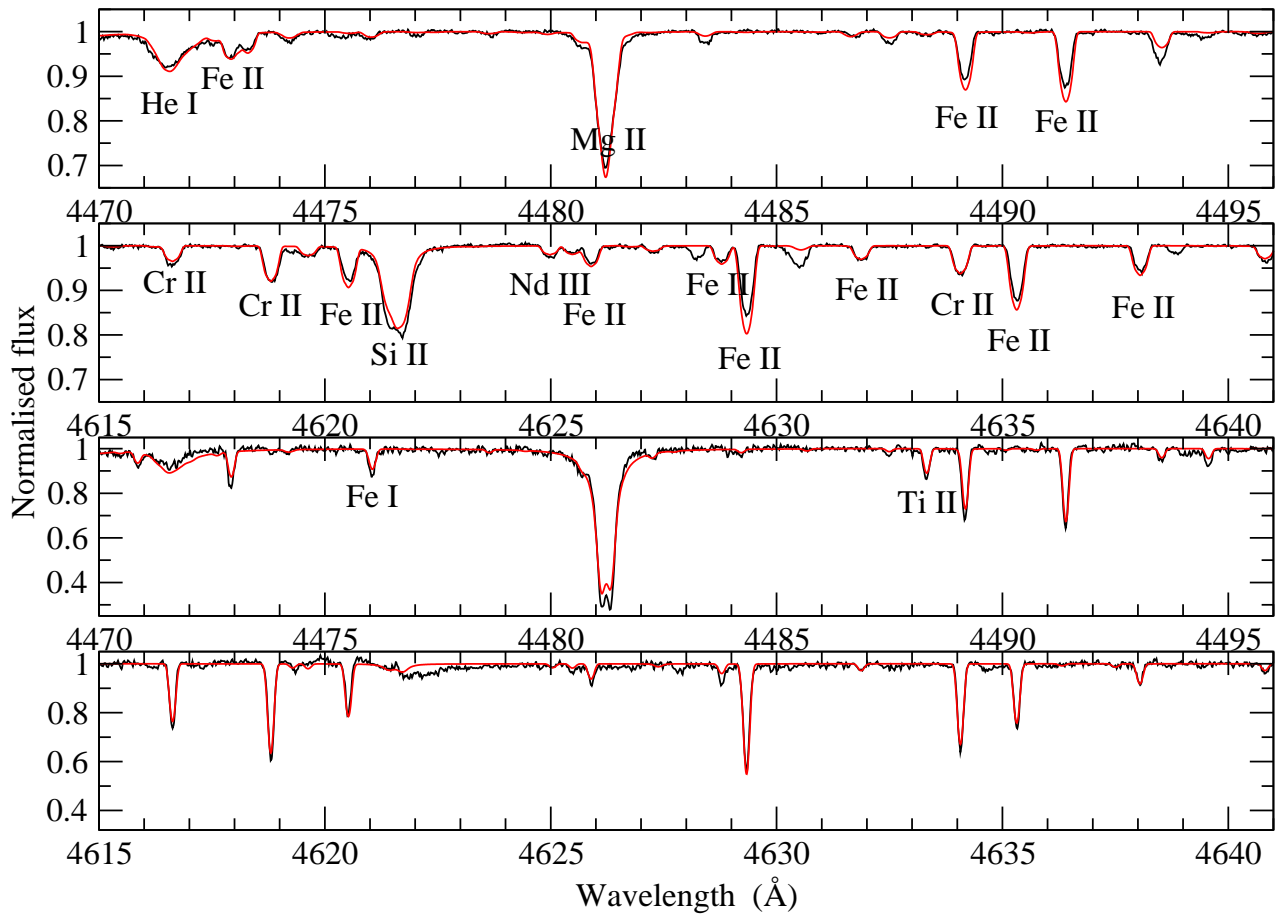}}
\caption{\label{Fig_spec_fit_A_B} Small sections of the best fits
found by {\sc zeeman.f} to the disentangled mean spectrum of \bd\,A
(upper two panes) and \bd\,B (lower two panes). The observed spectra
are the (black) curves with noise; the computed spectra (red) have no
noise. Ion identifications are provided for the stronger lines in the
upper two panes; these generally apply to the corresponding lines in
the lower two panes.  }
\end{figure*}

\begin{table}[th]
\begin{center}
  \caption{Properties of primary and secondary stars of \bd\ as
derived from this study}
\label{Tab_star_params}
\begin{tabular}{lll}\hline 
Property (units)         &   Primary (A)     & Secondary (B)   \\
\hline
Mass (M/$M_\odot$)       & $3.40 \pm 0.20$   & $2.30 \pm 0.15$  \\
Luminosity (\llo)        & $2.15 \pm 0.10$   & $1.50 \pm 0.10$  \\
Effective temperature \te (K)   & $13200 \pm 600$   & $10000 \pm 500$  \\
Radius ($R/R_\odot$)     & $2.27 \pm 0.36$   & $1.90 \pm 0.25$  \\
Gravity $\log g$ (cgs)   & $4.25 \pm 0.13$   & $4.26 \pm 0.13$  \\
Velocity \vsi\ (\kms)    & $13.8 \pm 0.4$    & $6.3 \pm 0.2$    \\
\hline
\end{tabular}
\end{center}
\end{table}

The Str\"{o}mgren $uvby$ colours of \bd\ were used to determine a
temperature, $\te = 12800$\,K, and luminosity, $\llo = 2.25$, of the
system treated as a single star. Using the single-star parameters, the
position of \bd\ in the theoretical HR Diagram lies essentially on the
isochrone for the cluster age. The position in the HR diagram provides
a single-star mass of $M = 3.5\,M_\odot$. The deduced value of the
fractional age (the fraction of main sequence lifetime already
elapsed) is 0.46 \citep{Landetal07}. (The isochrones used for HR
diagram comparisons are those of the Padova group, described by
\citet{Fagoetal94}, for $Z = 0.02$. The methods used to obtain
accurate cluster ages are described in detail by \citet{SilaLand14}).

Because the secondary star contributes to the spectrum and to the
photometric magnitudes, we anticipate that the primary is slightly
bluer than the observed colours, but about 0.1~dex less luminous (see
below), $\log(L_1/L_\odot) = 2.15 \pm 0.1$. The result is that the best
estimate of the mass of the primary changes only slightly, to about
$(3.40 \pm 0.2)\,M_\odot$. The estimated effective temperature of the
primary, from its position in the HR diagram, is about $\log \te =
4.12 \pm 0.02$, or $\te = 13200 \pm 500$\,K.  

Using the well-determined binary mass ratio of 1.479, we deduce that
the secondary star has a mass of $(2.30 \pm 0.15)M_\odot$. When we
place the secondary on the cluster isochrone, we deduce a luminosity
of about $\log (L_2/L_\odot) = 1.5 \pm 0.1$. This means
that the secondary contributes about 18\% of the total light from the
system, which is equivalent to the 0.1~dex assumed above. The
luminosity ratio is $L_1/L_2 \approx 4.5$. The secondary lies very close
to the cluster ZAMS, with $\te \approx 10000 \pm 500$\,K. 

Because the secondary contributes only a rather small fraction of the
light, the single star parameters are approximately representative of
the primary star of \bd, and the mass and other properties of the
primary are determined securely with reasonably good
precision. (Moreover, because the primary is a chemically peculiar
star, a more refined estimate of its parameters is not likely to be
significantly more accurate than our first simple estimate.) 
Similarly, the properties of the secondary are rather accurately
determined from the mass ratio and the fact that both stars are
members of the open cluster IC~4725 and thus have well-determined age,
distance and positions in the cluster HR diagram.

In order to investigate further the properties of the two stars in the
\bd\,AB system, we have carried out spectral disentangling using a
programme developed by one of us (O. Kochukhov). This code was
previously applied to SB2 systems containing CP star components by
\citet{Folsetal10,Folsetal13}. The reader can find more details about
our disentangling procedure in those publications.

The result of disentangling is two high S/N spectra,
one for each star, which when added together with appropriate radial
velocity shifts, match closely the (rather noisy) individual observed
binary spectra. However, the disentanglement itself does not provide
information about the fraction of the light contributed at each
wavelength by each of the two stars; each of the two solution spectra
is still normalised to the total light of the system at each
wavelength. 

For abundance determinations, we need to determine the flux ratio in
the wavelength region to be modelled (or equivalently the zero level
of each spectrum). Since component A of \bd\ is
significantly hotter than component B, and thus has substantially
higher flux in the UV, we suspect that the light ratio in the visible
window will probably be less than 4.5. In fact there is one spectral
window, around 4500\,\AA, where the local flux ratio is quite closely
constrained, by the depth of the Mg~{\sc ii} triplet of lines at
4481.13, 4481.15, and 4481.33\,\AA. The feature produced by this group
is so strong that the central depth of the line is set essentially (in
LTE) by the surface temperature at the top of atmosphere. Furthermore,
because the feature is a blend of lines about 0.2\,\AA\ apart, the
observed central depth of the line is almost independent of (small)
microturbulent velocity, and of rotational velocity provided that
\vsi\ is less than about 10\,\kms. In addition this central depth varies
little with \te\ for small excursions around $\te = 10000$\,K.

We have available high resolution spectra of spectral windows
including the Mg~{\sc ii} feature for several stars which have
parameters close to those deduced for \bd\,B, obtained using the \esp\
spectrograph ($R \approx 65000$) at the Canada-France-Hawaii Telescope
or the Aur\'{e}lie spectrograph ($R \approx 50000$) at the
Observatoire de Haute Provence. The available spectra include
HD\,114330 (minimum normalised flux = 0.270), HD\,72660 (0.260),
HD\,214994 (0.270), HD\,209459 (0.297), and HD\,19805 (0.267). All have
$\te = 9700 \pm 600$\,K, $3.4 \leq \log g \leq 4.3$, and $\vsi \leq
7$\,\kms.

From these observed values in similar stars, we adopt a typical
minimum normalised flux of $0.27 \pm 0.01$ for \bd\,B. When we
experiment with various values of flux ratio for normalising the two
disentangled spectra, this constraint on the central depth of the
Mg~{\sc ii} feature sets the zero point of the disentangled spectrum
of \bd\,B, which results in a flux ratio $f_1/f_2$ at
4481\,\AA\ of $3.6 \pm 0.1$. We assume that this ratio is valid over at
least $\sim 10^2$\,\AA\ in either direction from the Mg~{\sc ii} blend,
and normalise both disentangled spectra in this region appropriately. 

The disentangled, renormalised spectra around 4500\,\AA\ were then fit
using the line synthesis programme {\sc zeeman.f} \citep{Land88,
Wadeetal01}. This programme assumes a given value of \te\ and $\log
g$, interpolates within a grid of ATLAS solar abundance models, and
computes a synthetic spectrum using a line list from the Vienna Atomic
Line Database
\citep[VALD][]{Pisketal95,Ryabetal97,Kupketal99,Kupketal00}. The
computation is fully in LTE, and the effects of a magnetic field on
radiative transfer in each spectral line can be included (or omitted
if the star is non-magnetic). An initial abundance table is assumed,
and one element at a time is iterated to a best fit of all the largely
unblended lines of a single element found within the window under
study. This process can be iterated (for a non-magnetic atmosphere)
over various values of assumed microturbulence in order to select the
microturbulence parameter that leads to the best concordance between
abundances from weak and strong lines. At the end, the programme
reports the best fit abundance of the element varied, the best fit
value of \vsi, and the best microturbulent parameter $\xi$, and
produces a set of plottable synthetic spectra with the lines used for
the fitting marked. The process is then repeated for other elements,
and iterated to convergence.

We start spectrum fitting with \bd\,B, as the lack of a magnetic field
makes this the simpler of the two stars to model. Our fitting process
was first carried out for Fe in windows about 75\,\AA\ wide in
the range 4438 -- 4652\,\AA. Because iron has numerous lines in these
windows, some weak and others strong, our spectrum synthesis yields a
robust non-zero value of the microturbulence parameter of $\xi = 1.2
\pm 0.2$\,\kms. This value was held fixed for further calculations of other
elements for which, in general, not enough lines are available to
provide an accurate value of this parameter. The best fit value of
projected rotational velocity is found to be about $\vsi = 6.3 \pm
0.2$\,\kms, and the abundance of iron is found to be $\log N_{\rm
Fe}/N_{\rm H} = -4.32$, an overabundance of almost a factor of two
relative to solar abundance ratios. The dispersion of values suggests
an uncertainty in the derived iron abundance of about 0.1\,dex.

This fitting process was repeated for a window at
4998--5072\,\AA, where the normalisation of the two disentangled
spectra was assumed to still be given by $f_1/f_2 = 3.6$,
and the microturbulence was held at 1.2\,\kms. For this window a best
fit abundance of iron of $\log N_{\rm Fe}/N_{\rm H} = -4.15 \pm 0.1$
was found. This is not significantly different from the value found
around 4500\,\AA, and we conclude that the normalisation in this window
is not greatly different from that at 4500\,\AA, so we use the same
flux ratio of 3.6 for all spectra modelled in these two windows. 

Experiments varying the effective temperature \te\ between the
$1\sigma$ limits of 9500 and 10500\,K indicate that the abundance
uncertainty due to \te\ is about $\pm 0.20$~dex. This uncertainty
dominates the total error budget for iron. We find that the situation
is similar for other iron peak elements.

The fitting process was then repeated for the elements He, Mg, Si, Ca,
Ti, Cr, and Ba. The deduced abundance ratios are given in
Table~\ref{Tab_abunds_A_B}, together with the solar ratios. The final
fit of the synthetic spectrum to two small segments of the
renormalised mean spectrum of the secondary is shown in
Fig.~\ref{Fig_spec_fit_A_B}. It is clear that a reasonable fit has
been achieved for most of the lines in the illustrated spectrum
segment shown in the figure.

The underabundance of Ca relative to the solar abundance, and the
overabundance of Ba, strongly suggest that \bd\,B is a hot~Am
star. With $\te \approx 10000$\,K, it is near the upper temperature
limit of this class. 

For \bd\,A, which has a magnetic field, we need to make an assumption
about what field strength and geometry to include in the spectrum
synthesis. We first note that the dipolar field large enough to produce the
maximum mean longitudinal field absolute value observed, a little larger than
$400$~G, has a polar field strength below about 1500\,G. Since the
actual field configuration may depart significantly from that of a
simple dipolar field, the local field strength may exceed this value in
places on the stellar surface. Even allowing for this possibility, it
is reasonable to suppose that the field locally is generally 
below 2~kG. Such a weak field has very little effect on
the observed flux line profiles (about as large an effect as a
microturbulence of 1\,\kms). Thus we can derive an accurate value of $\vsi
= (13.8 \pm 0.4)$\,\kms\ by fitting line profiles of some of the stronger
lines, including a magnetic field of 1.5\,kG. 

Using the expression for the equatorial velocity of the primary star
$v_{\rm eq} = 50.6 (R/R_\odot)/P_{\rm rot}$ \citep[e.g.][]{Pres71}, together
with the inferred radius of the primary star, we find that $v_{\rm 1,eq}
= (22.8 \pm 3.6)$\,\kms, and with $v_1 \sin i_1 =
13.8$\,\kms, we deduce that the inclination of the rotation axis of
the primary star to the line of sight is $i_1 = 37^\circ \pm
9^\circ$. 

The magnetic field of the primary is expected to have a roughly
dipolar morphology. The observed $V/I$ signatures are generally rather
simple, like the one shown in Fig.~\ref{Fig_sample_lsd_plot},
consistent with this assumption. Assuming that the field can be
usefully modelled by a simple dipolar field, the angle $\beta$ of the dipole
axis with respect to the stellar rotation axis can be estimated using
Eqn.~(3) of \citet{Pres71} to be about $\beta = 26^\circ \pm
9^\circ$. That is, in order for the field moment \bz\ not to change
sign as the primary rotates, the field axis cannot make a large angle
with respect to the rotation axis. With the deduced angles, the line of
sight passes very close to the visible magnetic pole, and we can
estimate that the polar field strength of the model dipole is of the
order of 1400~G \citep[Eq.~(9)]{Pres71}. 

The disentangled spectrum of \bd\,A is an average over observed phases,
so we have taken the field to be a dipole plus octupole (chosen to reduce
the dipolar contrast between polar and equatorial field strengths)
with a typical strength of $\bs \sim 1200$~G in the fitting
process. The exact configuration of the field, and its exact value,
make very little difference to the deduced abundance
values. Proceeding as for the secondary star, we find the abundance
values listed in Table~\ref{Tab_abunds_A_B}. Again the uncertainties
in the abundance values include estimates of the scatter in values
deduced from different spectral windows, but are usually dominated by
the uncertainty in the value of \te. The derived abundances are
compared to solar abundance ratios from \citet{GrevSauv98}.

\begin{table}[ht]
\begin{center}
\caption{Atmospheric abundances of \bd\ A and B}
\label{Tab_abunds_A_B}
\begin{tabular}{lccc }\hline 
 Element & \multicolumn{3}{c}{$\log(N_{\rm X}/N_{\rm H})$} \\ & \bd\ A
 & \bd\ B & solar \\
\hline
 He      & $-2.48 \pm 0.20$ & $-0.92 \pm 0.30$  & $-1.07$ \\
 Mg      & $-5.84 \pm 0.10$ & $-4.68 \pm 0.10$  & $-4.42$ \\
 Si\,{\sc ii}  & $-3.68 \pm 0.20$ & $-4.08 \pm 0.10$ & $-4.45$ \\
 Si\,{\sc iii} & $-2.81 \pm 0.20$ & \dots            & $-4.45$ \\
 Ca      &  \dots           & $-6.20 \pm 0.20$ & $-5.64$ \\
 Ti      & $-7.15 \pm 0.20$ & $-6.95 \pm 0.20$ & $-6.98$ \\
 Cr      & $-5.95 \pm 0.20$ & $-5.87 \pm 0.20$ & $-6.33$ \\
 Fe      & $-4.10 \pm 0.20$ & $-4.32 \pm 0.20$ & $-4.50$ \\ 
 Ba      &   \dots          & $-8.86 \pm 0.40$ & $-9.87$\\
 Pr      & $< -8.6$         & \dots            & $-11.29$ \\
 Nd      & $-7.30 \pm 0.20$ & \dots            & $-10.50$ \\
\hline
\end{tabular}
\end{center}
\end{table}

Note that this star (like many other magnetic Bp stars) shows a large
discrepancy between the abundance values deduced from Si~{\sc ii} and
Si~{\sc iii}. We suspect that this arises from strong variation of the
abundance of Si with height in the atmosphere, as argued by
\citet{BailLand13}. 

\bd\,A is clearly strongly underabundant in He, by a factor of 20 or
more. Mg is also strongly underabundant. The iron peak elements Ti, Cr
and Fe are near solar or slightly overabundant. From the definite
identification of several weak lines, the rare earth Nd is clearly
present, and is overabundant by about 3.2~dex. These are abundance
values that clearly identify \bd\,A as a chemically peculiar (Bp) star,
consistent with its magnetic field. As expected for a star of mass of
about $3.4\,M_\odot$ that has an age of about $1.0\,10^8$\,yr and is
about half way through its main sequence evolution, the
root-mean-square magnetic field $\bz_{\rm rms}$ is below 1~kG
\citep{Landetal08}, and some of the chemical abundances are
approaching solar values, a result discovered for a sample of magnetic
Bp stars of similar mass by \citet{Bailetal14}. Note that Bailey et
al. included \bd\ in their study as a single star, finding abundances
in broad agreement with those found here, given the slightly different
fundamental parameters and the significantly different zero-point
choice adopted by them.

\section{The orbit again}

With estimates of the masses of the two stars in the \bd\ system, we
return to the geometry of the orbit. From the value of $M_1 \sin^3 i =
0.435\,M_\odot$, and $M_1 \approx (3.40 \pm 0.20)\,M_\odot$,
we deduce that the orbital inclination is given by $\sin i = 0.504 \pm
0.010$ or $i = 30.3^\circ \pm 0.6^\circ$. 

In turn the value of $i$ allows us to deduce (see
Table~\ref{Tab_sb2_params}) that $a_1 = 20.5\,R_\odot$ and $a_2 = 30.4
R_\odot$. The semi-major axis of the relative orbit is $a = a_1 + a_2
= 50.9\,R_\odot = 3.54\,10^{12}$\, cm. In the relative orbit of
the two stars, periastron occurs with a separation of $r_{\rm pa} =
(a_1 + a_2)(1-e) = 26.8\,R_\odot$ while apoastron occurs with a
separation of $r_{\rm aa} = (a_1 + a_2)(1 + e) = 75.0\,R_\odot$.

Using the parameters we have estimated in the previous section for the
two stars of \bd, we find that the radius of the primary is $R_1
\approx 2.3\,R_\odot$. Thus the semi-major axis of the relative orbit
is about $22.4\,R_1$ and the separation of the two stars at
periastron is about $r_{\rm pa} \approx 12\,R_1$. 

The two components of \bd\ are both stars with external radiative
envelopes and convective cores. For such stars, tidal friction is
driven by thermal diffusion applied on tidal gravity waves (the
so-called dynamical tide) excited in the radiative zones of the two
stars \citep{Zahn75}. For this situation, the ratio $a/R_1$ is
substantially above the critical value of the ratio required for
circularisation and synchronisation according to \citet{Zahn77}. Thus
it is not expected that this system should be able to reach a state of
a circularised orbit and synchronous stellar rotation during its main
sequence lifetime \citep[see also][]{NortZahn03}.

Although the separation of the components is too large to have
produced either orbital circularisation or rotational synchronisation,
the rather small minimum separation suggests that the two stars may
nevertheless have undergone some tidal interaction in the
$10^8$\,yr during which they have presumably orbited one another.

One consequence of tidal interaction (or of initial conditions during
formation of the binary) could be that the rotation axis
of the primary would be aligned to the orbital axis. If
this situation occurs, then the inclination angle of the stellar
rotation to the line of sight should be the same as that of the orbit,
$30^\circ \pm 1^\circ$. The derived inclination angle of the primary, $37^\circ
\pm 9^\circ$, is within the uncertainties the same as the orbital
inclination. Although the (near) equality of the two inclinations does
not prove that the primary star has its rotation axis normal to the
orbital plane, this result is consistent with that situation.

This computation strengthens the identification of 5.04\,d rather than
the 1.245\,d alias as the correct value of the rotation period of the
primary (see Sect.~4), since the 5\,d period leads to a ``reasonable''
value of $i_1$, while the shorter period would require $i_1 \approx
8^\circ$, a value which has no obvious relationship to the binary
orbit. 

Another consequence of the tidal interaction could be to alter or even
regulate the rotation period of the primary. Clearly the stellar
rotation is not synchronous with the mean orbital velocity. In fact,
the primary rotates with a period of only 5.04\,day, less than 1/3 of
the orbital period. Thus we are led to consider the possibility that
the primary star has its rotation period locked to the orbital motion,
not with respect to the average orbit period, but to the relative
motion of the secondary star at periastron, when the
tidal interaction is most intense. This possibility is related to the
``pseudo-synchronous states'' introduced and discussed by
\citet{Hut81}. We can test this idea by comparing
the rotational angular velocity of the primary star to the orbital
angular velocity of the secondary star at its closest approach to the
primary.

With a rotational period of $P_1 = 5.04$~d, the angular velocity of the
primary is $\Omega_1 = 2\pi /P_1 = 1.44\,10^{-5}$\,rad\,s$^{-1}$. 

The maximum velocity of the secondary star relative to the primary
star is given by 
\begin{equation}
     v_{\rm max}^2 = \frac{G(M_1 + M_2)}{(a_1 + a_2)} 
          \frac{(1+e)}{(1-e)}. 
\end{equation}
Using the estimates of the masses from the previous section, of
the orbital inclination $i = 30^\circ$, and of $e$, $a_1 \sin i$ and
$a_2 \sin i$ from Table~\ref{Tab_sb2_params}, we find that $v_{\rm
max} = 245$\,\kms. This is very close to the maximum observed velocity
difference between the two stars of about 123\,\kms divided by $\sin
i$, or 244\,\kms. This is what is expected from the value of the argument of
perihelion $\omega = 187^\circ$, which indicates that our line of
sight is almost perpendicular to the major axis of the relative orbit
of the two stars. From this result, the angular velocity of the
secondary star relative to the primary at periastron is given by
\begin{equation}
\Omega_{1-2} = \frac{v_{\rm max}}{(a_1 + a_2)(1-e)} = 1.34\,10^{-5}
     \,{\rm rad\,s^{-1}}, 
\end{equation}
which corresponds to a period of 5.4\,days.

\begin{table*}[t]
\begin{center}
  \caption{Properties of known short-period ($P_{\rm orb} < 20$\,d) SB2 systems containing a magnetic upper main sequence star }
\label{Tab_mag_sb2_sys}
\begin{tabular}{llllllllllllllllll}
\hline 
\multicolumn{2}{c}{System} & \multicolumn{3}{c}{Orbit} & \multicolumn{6}{c}{Primary}  & \multicolumn{3}{c}{Secondary}   \\
            &              & $P_{\rm orb}$ & $i$ & $e$ & $M_1$ & $P_1$ & $v_1 \sin i_1$ & $i_1$ & $B_{\rm d1}$ & $\beta_1$ & $M_2$ & $v_2 \sin i_2$ & $B_{\rm d2}$  \\
            &              & days   & deg &            & $M_\odot$ & days  & \kms     & deg   &  G    & deg       & $M_\odot$ & \kms     & G      \\
\hline
HD 5550     & HIP 4572     & 6.82   & \dots & 0.006    & 2.5:      & 6.84  & 5        & 32    &  80   & 24        & 1.7:      & 3        &  \\
HD 37017    & V1046 Ori    & 18.65  & \dots & 0.277    & 9:        & 0.901 & \dots    & 30    & 6500  & 50        & 7.3:      & \dots    & \dots \\
HD 37061    & NU Ori       & 19.139 & \dots & 0.14     & 13:       & \dots & 180      & \dots & 620   & \dots     & \dots     & \dots    & \dots \\
HD 47129    & Plaskett's   & 14.396 & 71  & 0          & 45.4      & 10.2  & 70       & \dots & \dots & \dots     & 47.3      & 305      & 2850 \\
HD 98088    & Abt's        & 5.905  & 69  & 0.184      & 2.19      & 5.905 & 23       &  77   & 3850  & 76        & 1.67      & 21       & 0    \\
HD 136504 & $\epsilon$ Lup & 4.56   & 21  & 0.277      & 8.7       & \dots & 37       &  21   & 900   & 0         & 7.3       & 27       & 600 \\
HD 161701   & HIP 87074    & 12.451 & 71  & 0.0043     & 4.0       & 12.451 & 17      & \dots & 0     & \dots     & 2.4       & 8        & $\geq 750$  \\
BD-19 5044L & IC 4725 98   & 17.630 & 30  & 0.474      & 3.4       & 5.04  & 14       & 37    & 1300  & 26        & 2.3       & 6        & 0    \\
\hline
\end{tabular}
\end{center}
\end{table*}

The two values of angular velocity are not equal but they are
remarkably similar. It does not appear that the uncertainty of
either angular velocity is large enough to make the two values
equal. However, their very similar values suggest that tidal
interaction at periastron may very well be affecting the rotation of
the primary, and that this effect is close to forcing the primary star
to rotate at the same angular rate as the orbital angular velocity of the
secondary star at periastron. 
\footnote{If we compute the pseudo-synchronous rotation following 
the formalism of \citet[][Eqs. 42 and 45]{Hut81}, we obtain a rotation
of 6.9 days for a constant time lag friction model. This shows that
the dynamical tide in intermediate and massive stars cannot be
modelled using such a simplified model.}

Another peculiarity of the primary star is that the inclination angle
$\beta$ of the magnetic field with respect to the stellar rotation
axis (see Sect.~5) is small, of the order of $26^\circ$. Since
magnetic Ap/Bp stars having rotation periods shorter than a few weeks
usually have large $\beta$ angles
\citep[e.g.][]{LandMath00}, we can wonder whether the small $\beta$
value is another consequence of tidal interactions. Interestingly, two
other close binary systems containing magnetic Ap/Bp stars, $\epsilon$~Lup
\citep{Shuletal15} and HD\,5550 \citep{Alecetal16}, also share the
characteristic of magnetic fields having small values of $\beta$.

\section{Conclusions}


During the course of a project to obtain repeated spectropolarimetric
observations of magnetic Ap/Bp stars in open clusters, we discovered
that the magnetic Bp star \bd, a member of the open cluster IC~4725,
is a short-period double-lined spectroscopic binary system. Because
short-period main sequence SB systems only very rarely contain a
magnetic Ap/Bp star, we have systematically observed this system to
determine the main characteristics of the individual stars composing
it, and the orbital parameters, and tried to identify ways in which the
binary nature may have interacted with the magnetic field of the
primary star. A particular interest of this system is that, in
addition to being an intrinsically rare type of close binary, its
membership in IC\,4725 independently tells us that its age is approximately
$1.0\,10^8$\,yr. 

The system is found to have an orbital period of $17.630 \pm
0.005$\,d. The orbit is quite eccentric, with $e = 0.474$. The mass
ratio is 1.479. The primary star can be securely identified as a late
B star of mass about $3.4\,M_\odot$ and $\te = 13200$\,K. The
secondary has a mass of about $2.3\,M_\odot$ and an effective
temperature of about 10000\,K. We have disentangled the two spectra
(obtaining mean spectra for both stars), and used spectrum synthesis
to carry out an approximate abundance analysis of each of the two
components. These results (Table~\ref{Tab_abunds_A_B}) show that
component A is clearly a magnetic Bp star with typical abundance
anomalies for its age, and that component B appears to be a
(non-magnetic) hot Am star.

With these data and a model of the double-lined binary velocity curve,
we determine that the orbit is inclined to the line of sight at $i =
30^\circ$, and that the semi-major axis of the relative orbit has a
length of $a = 3.58\,10^{12}$\,cm = $51.4\,R_\odot$. The closest
approach of the secondary to the primary occurs with a separation of
only about $12\,R_1$.

The magnetic field of the primary star is observed to vary
periodically with the stellar rotation period of 5.04~d. From the
observed variations we deduce that the rotation axis of the star is
probably perpendicular to the binary orbit, and that the field is
roughly dipolar, with an axis inclined to the rotation axis by roughly
$\beta = 26^\circ$. The field of the primary star has a polar field strength
of the order of 1400~G, while no field is detected in the secondary. 

The angular velocity of the primary star is almost, but not quite,
equal to the angular velocity with which the secondary orbits relative
to the primary at periastron, and angular momentum exchanges between
orbit and rotation may have led to this situation.

This SB2 system is of considerable interest in the context of stellar
magnetic fields. It is a new member of the tiny class of known close
binary stars containing a magnetic Ap/Bp star. Because the system is a
member of the open cluster IC~4725, the system age is known
independently. Unusually as well, enough is known about the basic
parameters of both stars to fully characterise the orbit of the
system, and to demonstrate that significant interactions of several
kinds may have occurred between the system members during $10^8$\,yr.

A table of all known short-period ($P_{\rm orb} < 20$\,d) SB2 systems
containing (at least) one magnetic star is shown in
Table~\ref{Tab_mag_sb2_sys}. For each star, the table lists the period
$P_{\rm orb}$, inclination $i$, and eccentricity $e$ of the orbit; the
mass $M_1$, rotation period $P_1$, projected rotation velocity $v_1
\sin i_1$, the inclination $i_1$ of the rotation axis of the star to
the line of sight, the estimated dipolar field strength $B_{\rm d1}$
of the field, and the obliquity of the dipole axis to the stellar
rotation axis of the primary star; and the mass, \vsi\ value, and
estimated dipole field strength of the secondary star. These data are
incomplete, and \dots\ indicate missing values. Many values are at
least somewhat uncertain; colons ``:'' indicate particularly uncertain
values.

The data in the table are derived mainly from the following
publications. The discovery of the field of HD\,5550, and the data in
the table, were reported by \citet{Alecetal16}. The field of HD\,37017
was discovered by \citet{BorrLand79} while most of the tabular data
were taken from \citet{Bohletal87}. The masses were estimated from
\te\ values. \citet{Petietal11} discovered the field of HD\,37061, and
the data are from that paper and from \citet{Abtetal91}. All data on
HD\,47129 are from the discovery paper by \citet{Grunetal13}. The field
of HD\,98088 was predicted by \citet{Abt53}, who observed Ap-like
spectrum variations, and confirmed directly by \citet{Babc58}; other
tabular data were taken from the substantial study by
\citet{Folsetal13}. A field was claimed on the basis of a single
marginal observation of HD\,136504 by \citet{Hubretal09} and clearly
detected by \citet{Shuletal12}; other data were taken from the paper
by \citet{Shuletal15} announcing the discovery of a field in the
secondary star of this system and analysing the system. The magentic
field of the secondary star in the HD\,161701 system was discovered by
\citet{Hubretal14}, and their analysis of this system was described in
more detail by \citet{Gonzetal14}. All data concerning \bd\ are from
the present paper.

In addition to the systems with periods shorter than 20~d that
are listed in Table~\ref{Tab_mag_sb2_sys}, several longer-period SB2
systems have high eccentricity and may possibly have undergone
significant tidal interactions. These include HD 1976
\citep{Neinetal14}, HD 34736 \citep{Semeetal14}, HD 36485
\citep{Petietal13}, HD 55719 \citep{Bons76}, and HD 135728
\citep{Freyetal08}. 

The most striking feature of Table~\ref{Tab_mag_sb2_sys} is its
shortness. It contains only three systems with orbital periods less
than about 10\,days, short enough that synchronised rotation is
expected and (mostly) observed, and even for these systems only one
orbit (and one of a system with $P_{\rm orb} = 12.45$\,d) have been
circularised by tidal friction
\citep[see][]{Zahn77}. A second feature is the wide range of primary
masses, from $2.2\,M_\odot$ to $45\,M_\odot$. In five of the systems,
the primary star is magnetic; in two systems (HD\,47129 and
HD\,161701), it is the secondary which is magnetic, and one system
($\epsilon$\,Lup) actually contains two magnetic stars. The typical
field strengths \bs\ (estimated as the polar field strength of a
dipolar field $B_{\rm d}$ large enough to lead to the observed values
of \bz), vary from less than one hundred to several thousand~G.

A third point of interest in connection in particular with \bd\ is the
fact that the system HD\,161720, with stars similar to those of \bd,
and a period of 12.45\,days, not much shorter than that of
\bd, has achieved a state of both circularisation of its orbit and
synchronism of the primary rotation with the orbital period. The
existence of this system strongly suggests that significant effects
of tidal interaction may be present in the \bd\,AB system. 

However, no obvious global regularities emerge immediately from
Table~\ref{Tab_mag_sb2_sys}. It appears that the main empirical
path to understanding the origin and evolution of such rare binary
stars will have to rely mainly on detailed studies of individual
systems, like the present study, until a larger sample of such
magnetic short period systems is available.

\begin{acknowledgements}

Work by J.D.L. and G.A.W. has been supported by the Natural Sciences and
Engineering Research Council of Canada. O.K. acknowledges financial
support from the Knut and Alice Wallenberg Foundation, the Swedish
Research Council, and the Swedish National Space
Board. S. M. acknowledges funding by the European Research Council
through ERC grant SPIRE 647383. S. M. thanks E. Bolmont for fruitful
discussions on pseudo-synchronous states.

\end{acknowledgements}

\bibliography{MyBiblio}

\end{document}